\documentstyle[12pt]{article}

\def\lsim{\raise0.3ex\hbox{$<$\kern-0.75em\raise-1.1ex\hbox{$\sim$}}}
\def\gsim{\raise0.3ex\hbox{$>$\kern-0.75em\raise-1.1ex\hbox{$\sim$}}}
\begin{document}
\begin{titlepage}
\begin{flushright}
{\bf HU-SEFT R 1995-08}\\
{\bf UCL-IPT-95-07}\\
{\bf CPT-95/PE.3185}
\end{flushright}
\vskip 3.0 cm
\begin{center}
{\Large\bf Top-Induced Electroweak Breaking in the Minimal Supersymmetric
Standard Model}
\vskip 2.0 cm
M. Chaichian\renewcommand{\thefootnote}{1}\footnote{Research Institute for
High Energy Physics,
P.O. Box 9 (Siltavuorenpenger 20 C) FIN-00014, University of Helsinki,
Finland}$^,$\renewcommand{\thefootnote}{2}\footnote{Laboratory of High Energy
Physics, Department of Physics, P.O. Box 9 (Siltavuorenpenger 20 C)
FIN-00014, University of Helsinki, Finland},
P. Chiappetta\renewcommand{\thefootnote}{3}\footnote{Centre de Physique
Th\'eorique, CNRS-Luminy F-13288 Marseille, Cedex 9, France},
J.-M. G\'erard\renewcommand{\thefootnote}{4}\footnote{Institut de Physique
Th\'eorique, Universite Catholique de Louvain, B-1348 Louvain-la-Neuve,
Belgium}, R. Gonzalez Felipe$^1$ and J. Weyers$^4$
\vskip 4.0 cm
{\bf Abstract}
\end{center}

Severe constraints on parameters of the minimal supersymmetric standard
model follow from a dynamical electroweak symmetry breaking mechanism
dominated by top and stop loops. In particular, the lightest Higgs boson mass
is expected to be smaller than 100 GeV.
\end{titlepage}

{\Large\bf Introduction}
\vskip 0.5 cm
The recent confirmation \cite{CDF} of the experimental discovery of the top
quark with
a mass $m_t\approx$ 175-200 GeV is yet another success of the standard model
(SM) of
electroweak interactions. Indeed, theoretical computations of radiative
corrections to the SM confronted with the precise LEP data led to a range
of values for $m_t$ in remarkable agreement with the experimental results. To
appreciate the significance of these theoretical and experimental achievements,
it is worth recalling that calculations and measurements of the $g-2$ factor
of the electron do not allow for a determination of the muon mass with
anything even close to the precision of the top mass prediction in the SM !

Even though the SM is in remarkable agreement with all experimental facts,
it is fair to say that the origin as well as the hierarchy of quark and
lepton masses remain poorly understood. These issues are related to the
spontaneous symmetry breaking (SSB) of the electroweak gauge symmetry. The
usual tree-level Higgs mechanism for SSB is unsatisfactory in many ways and one
of the most attractive alternatives suggested so far is due to Coleman and
Weinberg \cite{coleman} (see also \cite{weinberg}): in their approach, SSB
is a dynamical effect induced  by radiative
corrections and not an ad-hoc input at the tree level of the theory.

Whatever the physics underlying SSB, the discovery of a Higgs scalar
\cite{gunion} is of
course a tantalizing challenge to experimentalists. On the other hand it is
clearly interesting and important to determine theoretically possible ranges
of values i.e., bounds for the mass $m_h$ of the Higgs in the SM and
of the lightest scalar in
as wide a class of models extending the SM as possible.

In the context of the SM viewed as an effective theory, i.e. valid up to some
scale $\Lambda$, the usual Higgs mechanism leads to a lower bound on $m_h$
determined from vacuum stability arguments \cite{altarelli}, while an
upper bound follows from
the requirement that no Landau pole appears up to the scale $\Lambda$
\cite{sher}.

In a recent letter \cite{fatelo}, very much inspired by the Coleman-Weinberg
approach, it was
suggested that the breaking of the electroweak gauge symmetry in the SM is a
quantum dynamical effect driven by top quark loops. The physical motivation
for this breaking mechanism comes from the fact that, with $m_t\simeq$ 200 GeV,
the Yukawa coupling of the top quark to the Higgs scalar is of order one and
thus significantly larger than all other electroweak coupling constants. A
straightforward consequence of this SSB mechanism is an upper bound on $m_h$
of the order of 400 GeV. It is however tempting to speculate further that top
loops are responsible for the physical effects of SSB, i.e. that the full
``Higgs
potential" is dynamically generated by quantum effects. The consistency of
this point of view requires the effective scalar quartic interaction to be
negligibly small if not zero, indicating a possible connection with the
``triviality" of a $\lambda\phi^4$ theory \cite{fernandez}. Be that as it
may, the
phenomenological implementation of this additional assumption leads to a
cut-off $\Lambda$ of the order of 1 TeV and to a Higgs mass $m_h\approx$ 80
GeV.

The purpose of the present paper is to explore a bit further the
phenomenological implications of such a dynamical symmetry breaking mechanism
in the context of the minimal supersymmetric extension of the standard model
(MSSM) where the scalar quartic interactions are functions of the
electroweak couplings and, therefore, naturally small.

The attractive features of MSSM are well known \cite{nilles}. Among others, it
provides a
plausible solution to some of the puzzles inherent to the SM like the ``gauge
hierarchy" problem. Furthermore, MSSM fits nicely with unification ideas
since it is naturally embedded in many unified
supergravity and superstring models. If there is new physics beyond the SM,
MSSM is certainly one of the prime candidate theories to describe it.

In MSSM there are two $SU(2)_L\otimes U(1)$ Higgs doublets and, hence, in the
end there will be three neutral Higgs scalars (two CP-even ones and one
CP-odd one). For simplicity we will view MSSM as an effective theory with three
characteristic scales : the cut-off $\Lambda$ (which, as it will appear,
plays almost no role in our arguments\renewcommand{\thefootnote}{*}\footnote{
Because of this weak dependence of
the results on the cut off $\Lambda$, a renormalized version of the theory
would lead to similar conclusions.}), the global supersymmetry breaking
scale $M_S$ and, finally, the electroweak breaking scale given by a vacuum
expectation value (VEV) $v$. Our line of argument starts as usual in the
context of
MSSM. First, supersymmetry is broken. As a result all SUSY partners
of the
SM particles become heavy with masses of the  order of $M_S$. In the Higgs
sector, the CP-odd scalar and one linear combination, characterized by a
mixing angle $\beta$, of the two CP-even scalars become heavy with a mass
again of the order of $M_S$, while the orthogonal CP-even combination remains
light (with a positive mass). At this stage $SU(2)_L\times U(1)$ is still
an exact gauge symmetry of the model.

We proceed by assuming that SSB of the gauge group is completely driven
by top and stop loops since their couplings to the light scalar field are
equal and large. Consistency of this picture of SSB requires the quartic
scalar self coupling to be small. Putting it equal to zero leads to our final
results
$$
m_h \ \lsim\ 100\ {\rm GeV},\ \ M_S \ \lsim\ 600\ {\rm GeV} \  {\rm and} \ \
0.6 \ \lsim \tan \beta \ \lsim \ 1.6 \ .
$$

The paper is organized as follows : in Section I we briefly review well
known features of the Higgs sector in MSSM and in Section II we derive
and present our results. We then conclude with some brief comments on these
results as well as on earlier investigations concerning the  Higgs masses
in MSSM.
\vskip 1.0 cm
\section{The Higgs sector in MSSM}
Let us recall \cite{gunion} that in MSSM there are two complex
scalar $SU(2)_L$ doublets
of opposite hypercharge: $H_1=\left(\begin{array}{l}
H^0_1\\
H^-_1 \end{array}\right)$ which couples to the charged leptons and to
down-type quarks and $H_2=\left(\begin{array}{l}
H^+_2\\
H^0_2 \end{array}\right)$ which couples to up-type quarks only.

The tree level scalar potential for the real components of the CP-even
neutral scalars $h_1\equiv Re\,H^0_1$ and $h_2\equiv Re\,H^0_2$ reads
$$
V_0=\frac{1}{2}m^2_1h^2_1+\frac{1}{2}m^2_2h^2_2-m^2_{12}h_1h_2+
\frac{\tilde{g}^2}{32}(h^2_2-h^2_1)^2\ ,
\eqno{(1)}
$$
where $\tilde{g}^2=(g^2_1+g^2_2)$ and $g_{1,2}$ are the usual $U(1)$ and
$SU(2)$ gauge couplings.

Generically, the mass terms in Eq. (1) break supersymmetry softly. For our
purposes we require the potential $V_0$ to be bounded from below, i.e.
$$
m^2_1+m^2_2 \geq 2m^2_{12}
$$
and to be such that the gauge symmetry $SU(2)_L\otimes U(1)$
remains unbroken at the tree level:
$$
m^2_1m^2_2>m^4_{12} \ .
$$

As usual in this context, we assume that all SUSY partners of SM
particles have become heavy with masses of the order of the global SUSY
breaking scale $M_S$ and that one linear combination of the neutral
scalars
$$
h=h_1\cos \beta +h_2\sin \beta
\eqno{(2)}
$$
remains light, while the orthogonal combination
$$
H=-h_1\sin\beta +h_2\cos\beta
\eqno{(3)}
$$
has a mass of the order of $M_S$.

The relevant part of the tree level potential for the field $h$ then reads
$$
V_0=\frac{1}{2}\mu^2h^2+\frac{1}{32}\tilde{g}^2\cos^2 2\beta \  h^4
\eqno{(4)}
$$
with $\mu^2>0$ and the mixing angle $\beta$, defined by Eqs. (2)-(3) is
determined by the mass parameters in Eq. (1).

We insist on the fact that at this stage only supersymmetry has
been broken: $h$ and $H$ are normal scalar fields with positive masses and no
vacuum expectation values. We notice that many supergravity inspired models
imply the boundary condition $m^2_1=m^2_2$ with $m^2_{12}\neq 0$ around
the GUT scale. In this case $m^2_H\approx m^2_1+m^2_{12}\ ,\ \mu^2\approx
m^2_2-m^2_{12}$, and $\beta=\pm \pi/4$. It is interesting to note that the
latter relation $\beta=\pm\pi/4$ would also emerge from the ``triviality"
of the scalar potential $V_0$, i.e. from the requirement $\partial^4
V_0/\partial h^4=0$.

In Eq. (4) we have omitted terms of the form $h^2H^2,hH^3$ and $h^3H$.
It turns out that the electroweak gauge boson contributions to the one loop
effective potential are negligible because of the relative smallness of the
gauge couplings compared to the top Yukawa coupling. We expect the same
suppression to occur for $H$ virtual contributions since its couplings
to $h$ are
also given by the gauge couplings. In addition $m_H$ could also be assumed
to be much larger than $M_S$.
\vskip 1.0 cm
\section{The effective potential at one loop}
We now proceed in a completely standard fashion with the computation of the
dominant one-loop radiative corrections $V_1$ to the scalar potential $V_0$.
In other words we compute the infinite series of diagrams given in
Fig. $1a$ with massless top quarks in the loop, and in Fig. $1b$
with left and right stops in the loop. We assume that the stops have the
same soft
mass $m_s$ (of the order of $M_S$) and neglect possible left-right mixing
effects\renewcommand{\thefootnote}{**}\footnote{Notice that after the breaking
of the gauge symmetry when the top quark has acquired a mass $m_t$, the mass
of both
left and right stops will be given by $m^2_{\tilde{t}} \simeq m^2_s+m^2_t$,
when their mixing is neglected.}.

The resulting effective potential is well known and reads
$$
V=V_0+V_1
$$
with $V_0$ given by Eq. (4) and
$$
V_1=\frac{N_c}{8\pi^2}\int\limits^{\Lambda^2}_0 dk^2\,k^2\ln
\left(\frac{k^2+m^2_s+
g^2_th^2/2}{k^2+g^2_th^2/2}\;\frac{k^2}{k^2+m^2_s}\right)\ ,
\eqno{(5)}
$$
$N_c$ is the number of colours, $\Lambda$ is the cut-off scale and
$g_t=h_t\sin\beta$ is the top coupling to the scalar $h;\;h_t$ is the
supersymmetric Yukawa coupling of the top to $H_2$ and the factor
$\sin\beta$ follows from Eq. (2).

Note that when $m_s=0\ ,\ V_1=0$ as it should, while for $m_s\rightarrow
\infty$ we recover the top loop radiative corrections to the scalar potential
in an effective SM.

Minimization of the potential $V$ now yields a non-trivial minimum $<h>\equiv
v\neq 0$ : top and stop loops effects have thus induced the SSB of
the electroweak gauge group. At $<h>=v$, one obtains the equation
$$
\mu^2+\frac{1}{2}m^2_Z\cos^2
2\beta+\frac{N_c}{4\pi^2}\;\frac{m^2_t}{v^2}\left\{
m^2_t\ln\left(1+\frac{\Lambda^2}{m^2_t}\right)-(m^2_t+m^2_s)\ln\left(1+
\frac{\Lambda^2}{m^2_t+m^2_s}\right)\right\}=0\ ,
\eqno{(6)}
$$
where
$$
m^2_Z=\frac{1}{4}\tilde{g}^2v^2 \ , \ (v=246\ {\rm GeV})
\eqno{(7{\rm a})}
$$
and
$$
m^2_t=\frac{1}{2}g^2_tv^2\ .
\eqno{(7{\rm b})}
$$

Since $\mu^2$ is assumed positive, the expression in brackets in
Eq. (6) must be negative. This is always the case for any  $m^2_s > 0$.
We postpone further comments on Eq.(6) to the next section.

For now, let us compute the lightest Higgs boson mass at the scale $v$ i.e.
$m^2_h=(\partial^2V/\partial h^2)|_{h=v}$. Using Eq. (6), this yields
the result
$$
m^2_h=m^2_Z\cos^2 2\beta+\frac{N_c}{2\pi^2}\;\frac{m^4_t}{v^2}\left\{\ln
\left[\left(1+\frac{m^2_s}{m^2_t}\right)\left(\frac{\Lambda^2+m^2_t}
{\Lambda^2+m^2_t+m^2_s}\right)\right] \right. $$
$$
\left.-\frac{\Lambda^2m^2_s}{(\Lambda^2+m^2_t)(\Lambda^2+
m^2_t+m^2_s)}\right\}\ .
\eqno{(8)}
$$
Notice that $m^2_h$ is weakly cut-off dependent for $\Lambda\ \gsim$ 1 TeV and
remains finite in the limit $\Lambda\rightarrow\infty$.

We thus obtain the upper bound
$$
m^2_h\leq m^2_Z\cos^2 2\beta +\frac{N_c}{2\pi^2}\;\frac{m^4_t}{v^2}\ln
\left(1+\frac{m^2_s}{m^2_t}\right)\ .
\eqno{(9)}
$$

Eqs. (8) and (9) are our first results. They follow directly from the
assumption that top and stops loops dominate the radiative corrections to
the scalar potential and drive the SSB of the gauge group. In Fig. 2
we plot the upper limit on the light Higgs mass as given by Eq. (9) for
two extreme values of the mixing angle $\beta$.

For our physical assumptions to be coherent one should argue further that
the $h$ scalar self interactions remain negligible, at least at the
scale $v$. This is easily implemented phenomenologically by the condition

$$
\frac{\partial^4V}{\partial h^4}\left| _{h=v}\simeq 0\ \right. \ .
\eqno{(10)}
$$

Taken as an exact equation, Eq. (10) looks like a ``triviality
condition" on a scalar field theory. Again we postpone comments on this
point to the next section but for the moment we consider Eq. (10) as a
phenomenological constraint on the effective potential which is important
to guarantee, a posteriori, that $h$ scalar loops will not affect
the SSB mechanism induced by top and stop loops. Putting it
differently, if the physics of SSB is indeed dominated by radiative
corrections due to top and stop loops then scalar self interactions must
be small, at least at the scale where SSB occurs.

After some simple algebra, Eq. (10) takes the form
$$
\frac{1}{4}\tilde{g}^2\cos^2 2\beta+\frac{N_c}{2\pi^2}\;\frac{m^4_t}{v^4}
\{f(x,y)-f(0,y)\} \cong 0\ ,
\eqno{(11)}
$$
where
$$x=\frac{m^2_s}{m^2_t}\ ,\ y=\frac{\Lambda^2}{m^2_t}
\eqno{(12)}
$$
and
$$
f(x,y)=-\ln (1+\frac{y}{1+x})-\frac{9+x}{1+x+y}+\frac{4(2+x)}{(1+x+y)^2}$$
$$
-\frac{8}{3}\frac{1+x}{(1+x+y)^3}-\frac{4x(1+2x)}{3(1+x)^2}\ .
\eqno{(13)}
$$
We now proceed to discuss the phenomenological content of Eqs. (8) and (11).

For simplicity let us start with the limiting case (of large cut off)
$y\rightarrow\infty$. For Eq.(11) to admit a solution we must then demand
that
$$
\lim\limits_{y\rightarrow\infty}\,[f(x,y)-f(0,y)] \ \lsim \ 0
$$
or, explicitly, that
$$
F(x)\equiv\ln (1+x)-\frac{4x(1+2x)}{3(1+x)^2}\ \lsim \ 0\ .
\eqno{(14)}
$$
The function $F(x)$ is plotted in Fig. 3. The equation $F(x)=0$ has a
unique solution, namely $x_0=8.65$, which is the maximum value of $x$ allowed
by Eq. (11). This requires
$$
m_s\ \lsim \ 2.9\ m_t\ .
\eqno{(15)}
$$
On the other hand, the function $F(x)$ reaches its minimum value for
$x\simeq 2.16$ and $F(2.16)\simeq -0.38$ . This immediately leads to an
upper bound for $\cos^2 2\beta$ or to a range of values for $\tan \beta$
given by
$$
\left(\frac{1-0.32g^2_t}{1+0.32 g^2_t}\right)^{1/2}\lsim \ \tan \beta \ \lsim
\left(\frac{1+0.32 g^2_t}{1-0.32 g^2_t}\right)^{1/2}\ .
\eqno{(16)}
$$
The solution of Eq. (11) in the limit $\Lambda\rightarrow \infty$ is
expressed in
Fig. 4  as a plot of $\tan \beta$ in terms of $m_s$ for the central values
of $m_t$ in the two Fermilab experiments, namely $m_t=176$ GeV and
$m_t=199$ GeV.

Eq. (11) thus provides two important constraints on the parameters of MSSM :
an upper limit on $m_s$ (Eq. (15)) and a range of values for $\tan \beta$
(Eq. (16)).

Provided $\tan \beta\neq 1$, we can solve Eq. (11) for $\cos^2 2\beta$ and
substitute the result into Eq. (8) to obtain
$$
m_h=\frac{2}{\pi}\;\frac{m^2_t}{v}\left\{\frac{y^3}{(1+y)^3}-\frac{1}{2}
\frac{(3x+2y+3x^2+3xy)y^2}{(1+x)^2(1+x+y)^3}\right\}^{1/2}\ .
\eqno{(17)}
$$
Notice that in the limit $x\rightarrow\infty$, Eq. (17) reproduces the
result found in Ref. \cite{fatelo}, namely
$$
m_h=\frac{2}{\pi}\;\frac{m^2_t}{v}\left\{1+O\left(\frac{1}{y}\right)
\right\}\ .
\eqno{(18)}
$$

In the limit $y\rightarrow\infty$ Eq. (17) becomes
$$
m_h=\frac{2}{\pi}\;\frac{m^2_t}{v}\left\{1-\frac{1}{2}\;\frac{(2+3x)}{(1+x)^2}
\right\}^{1/2}\ ,
\eqno{(19)}
$$
meaning, in particular, that $m_s$ is a natural cut off for the effective
standard model considered in Ref. \cite{fatelo} if $x \gg 1$.

Eq. (19) thus gives an upper bound on $m_h$ in the limit $\Lambda\rightarrow
\infty$. To illustrate the weak sensitivity of our results on the cut-off
scale $\Lambda$,  we have plotted in Fig. 5 $m_h$  (as given by Eq. (17)) as
a function of $m_s$ for $\Lambda =1$ TeV and $\Lambda=\infty$.

Finally, let us briefly comment on our results in the particular case
$\beta=\pm \pi/4$. The Higgs mass is then given by the second term in Eq. (8),
namely
$$
m^2_h=\frac{N_c}{2\pi^2}\;\frac{m^4_t}{v^2}\left\{\ln \frac{(1+x)(1+y)}{1+x+y}-
\frac{xy}{(1+y)(1+x+y)}\right\}\ ,
\eqno{(20)}
$$
where $x$ and $y$ now satisfy the equation $f(x,y)=f(0,y)$ with $f(x,y)$
as in Eq. (13). For $\Lambda \rightarrow \infty$ and $m_s \simeq 2.9 \ m_t$,
one finds $m_h \simeq 74$ GeV (resp. 94 GeV) for $m_t = 176$ GeV (resp. 199
 GeV), as illustrated in Fig. 5. This concludes the presentation
of our results. Clearly the most striking phenomenological implications
of our model for SSB of the electroweak gauge group in the context of MSSM
are the ones already mentioned in the introduction, namely
$$
m_h \ \lsim\ 100\ {\rm GeV}\ ,
\eqno{(21)}
$$
$$
m_s \ \lsim\ 600\ {\rm GeV}
\eqno{(22)}
$$
and
$$
0.6 \ \lsim \tan \beta \ \lsim \ 1.6\ .
\eqno{(23)}
$$
\vskip 1.0 cm
\section{Comments}
The basic physical idea of the present paper is that in MSSM, spontaneous
symmetry breaking of the electroweak gauge group is a quantum effect
predominantly driven by top and stop loops. Implementation of this SSB
scenario puts very severe constraints on the allowed range of values for
some of the most important phenomenological parameters in MSSM. These
constraints are summarized in Eqs. (21)-(23) and imply in particular, a
very light Higgs scalar which should be observable \cite{sopczak} at LEP 200
if not before and a rather low supersymmetry breaking scale.

To conclude, we present a series of miscellaneous comments which among
others express our views on the reliability of the calculations presented
here as well as on the relation of our work to other approaches to SSB
in MSSM.

1. In the effective potential, we have neglected $W$ and $Z$
boson loops as well as the corresponding gauginos loops. An explicit
calculation of their effect, say on the Higgs boson mass, shows indeed
that it is at most of the order of one percent. The main reason for this is of
course that the gauge couplings $g_1$ and $g_2$ are significantly smaller
that $g_t$ - mutatis mutandis we expect the same argument to hold for $H$ and
the corresponding higgsino loops . Alternatively, one can always push the
mass of $H$ and its SUSY partner $\tilde{H}$ high enough so as to make their
effect on $V_1$ completely negligible.

2. It should be noted that $\beta$ is originally a mixing
angle defined by the diagonalization of a mass matrix (Eq. (1)). At the one
loop level, the same $\beta$ (with small corrections) is related to the vacuum
expectation values $v_{1,2}$ of the fields $H^0_{1,2}$. Indeed, since
$<H>=0$ we obtain $\tan \beta=v_2/v_1$, as usual. The VEV of $h$ is
of course given by $v=(v^2_1+v^2_2)^{1/2}$.

3. On several occasions we have referred to the ``triviality issue"
in a $\lambda\phi^4$ theory. Let us recall that in a pure $\lambda\phi^4$
theory the renormalized coupling constant $\lambda_R$ must be zero or else
the theory blows up at some scale (related to the value of $\lambda$). The
argument is essentially non perturbative \cite{fernandez}. We are not
suggesting that
triviality should be imposed order by order in the loop expansion of the
effective potential although such a procedure probably deserves further
study. For the purposes of this paper, it is quite sufficient to take the
more phenomenological point of view emphasized in section II : to insure
that top and stop loops do indeed dominate the effective potential the
quartic scalar self coupling must be small, at least at the scale $v$.

4. The results presented in section II are weakly cut-off
dependent. For a very large cut off we expect QCD corrections to our
results to become huge. For $\Lambda$ around 1 TeV on the other hand,
QCD corrections remain small.

5. Two sine qua non conditions for our approach to SSB to
make sense is that higher loop corrections to the one loop effective potential
remain small and that the ``one loop vacuum" is stable. As usual it is easy
to argue that higher loop effects will indeed be negligible \cite{coleman}
but vacuum
stability is a much more difficult condition to analyze. We have no insight
on this problem.

6. Of course many interesting papers have already been published
on Higgs mass predictions or bounds in MSSM. They can be roughly classified
according to the value of $\mu^2$ : in the usual scenario $\mu^2$ is
supposed to be negative at the weak scale \cite{okada} while the critical
case $\mu^2=0$
has also been considered \cite{diaz}. In both approaches, it has been
emphasized that a heavy top leads to large corrections to the tree level
bound or prediction for $m_h$.

In the case $\mu^2<0$, $m_h$ is expected to be larger than 120 GeV while for
$\mu^2=0,\;m_h$ is around 100 GeV. It is thus not surprising that in a
scenario with $\mu^2>0$, $m_h$ is predicted to be even lighter.
\vskip 1.0 cm
\noindent
{\large\bf Acknowledgements}
\vskip 0.5 cm
We are grateful to K. Huitu and A. Kupiainen for useful discussions.
One of us (P.C.) would
like to thank the Research Institute for High Energy Physics (SEFT),
University of Helsinki for the hospitality. This work was partially supported
by the EEC Science Project SC1-CT91-0729.

\newpage
\noindent
{\large\bf Figure Captions}
\vskip 1 cm
\noindent
{\bf Fig. 1} \ One-loop radiative corrections to the scalar potential $V_0$
due to top quark and stop contributions.  The infinite series of diagrams in
Fig. $1a$ represents the massless top quark contribution to the one-loop
effective potential, while the one in Fig. $1b$ corresponds to left and right
stops with soft mass $m_s$.
\vskip 1 cm
\noindent
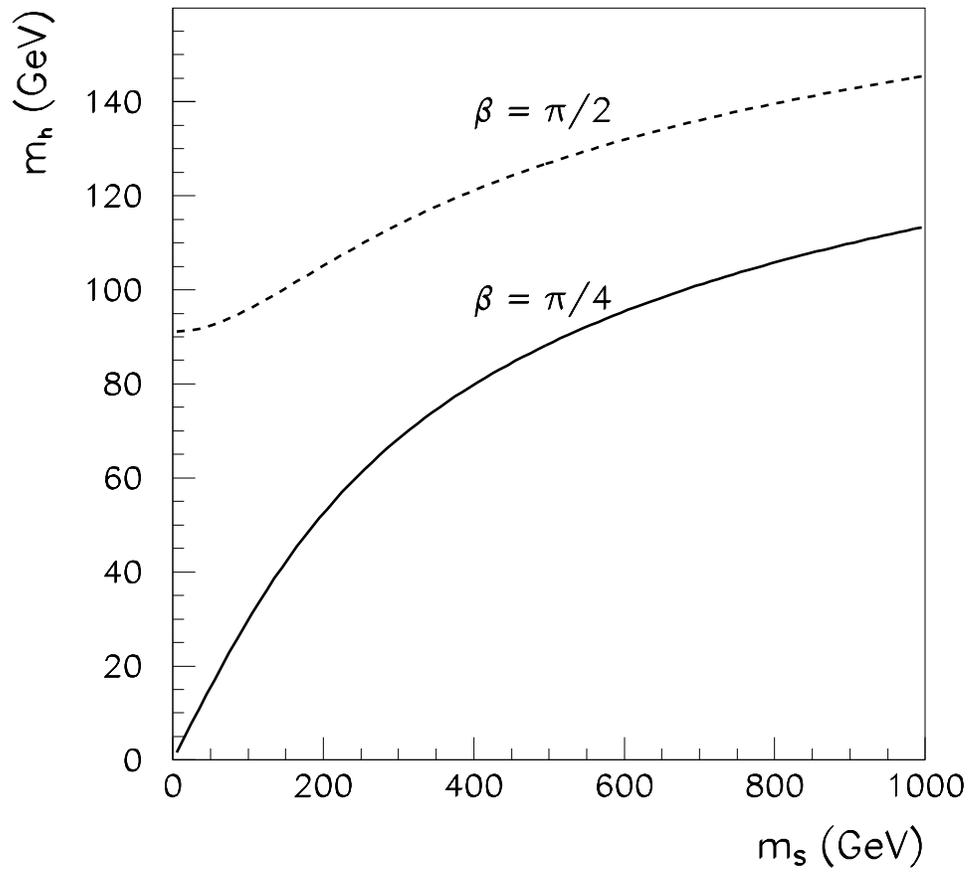
{\bf Fig. 2} \ Upper limit on the Higgs mass (Eq. (9)) under the assumption
that top
and stops loops dominate the radiative corrections to the scalar potential and
drive the SSB of the gauge group. The plot is given for the
extreme values of the mixing angle $\beta=\pi/2$ and $\beta=\pi/4$ and for
$m_t < 200$ GeV.
\vskip 1 cm
\noindent
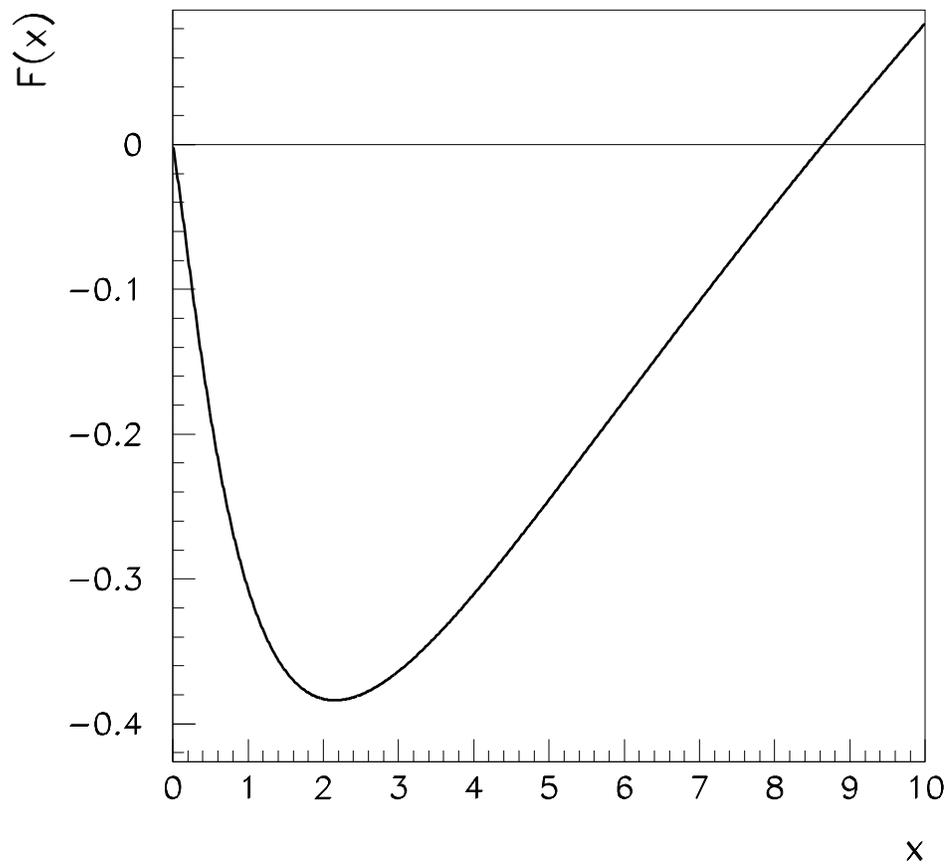
{\bf Fig. 3} \ Function $F(x)$ as given by Eq. (14) ($x=m^2_s/m^2_t$). The
solution of the equation $F(x)=0$ gives an upper bound on the soft mass
parameter $m_s$, $m_s \ \lsim \ 2.9 \ m_t$.
\vskip 1 cm
\noindent
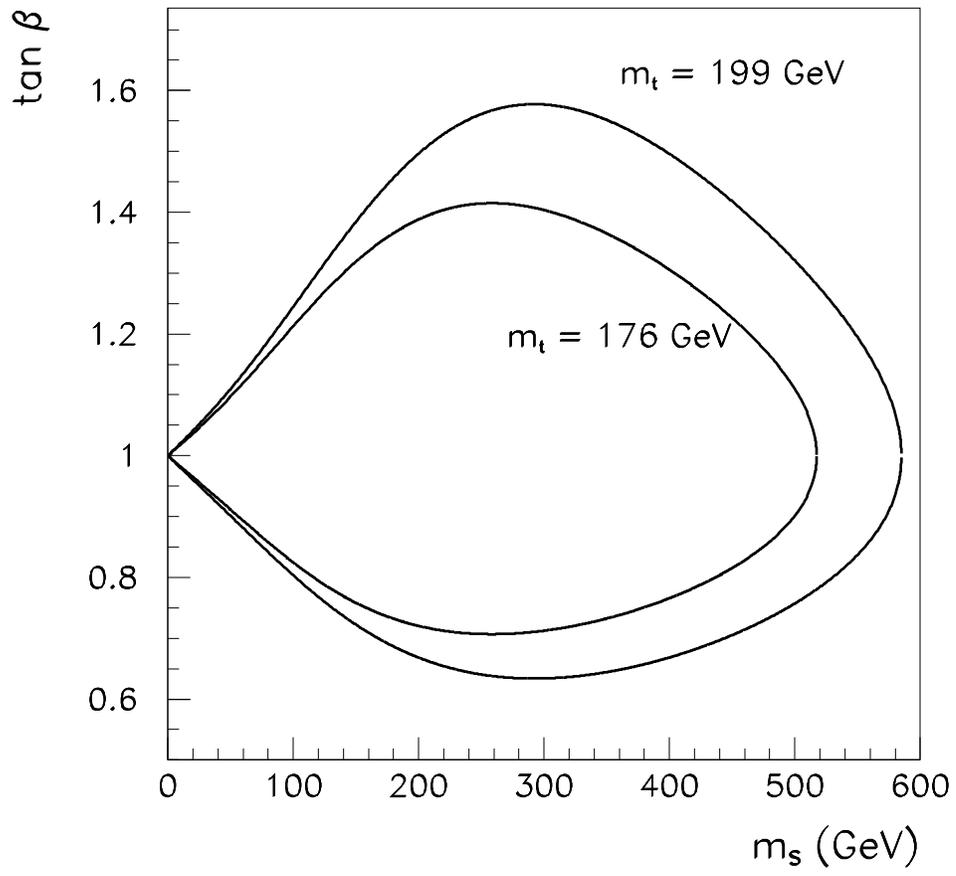
{\bf Fig. 4} \ Allowed range of values of $\tan \beta$ as a function of $m_s$
in the limit $\Lambda \rightarrow \infty$ for the central values of $m_t$
reported by the two Fermilab experiments,
namely $m_t = 176$ GeV and $m_t = 199$ GeV.
\vskip 1 cm
\noindent
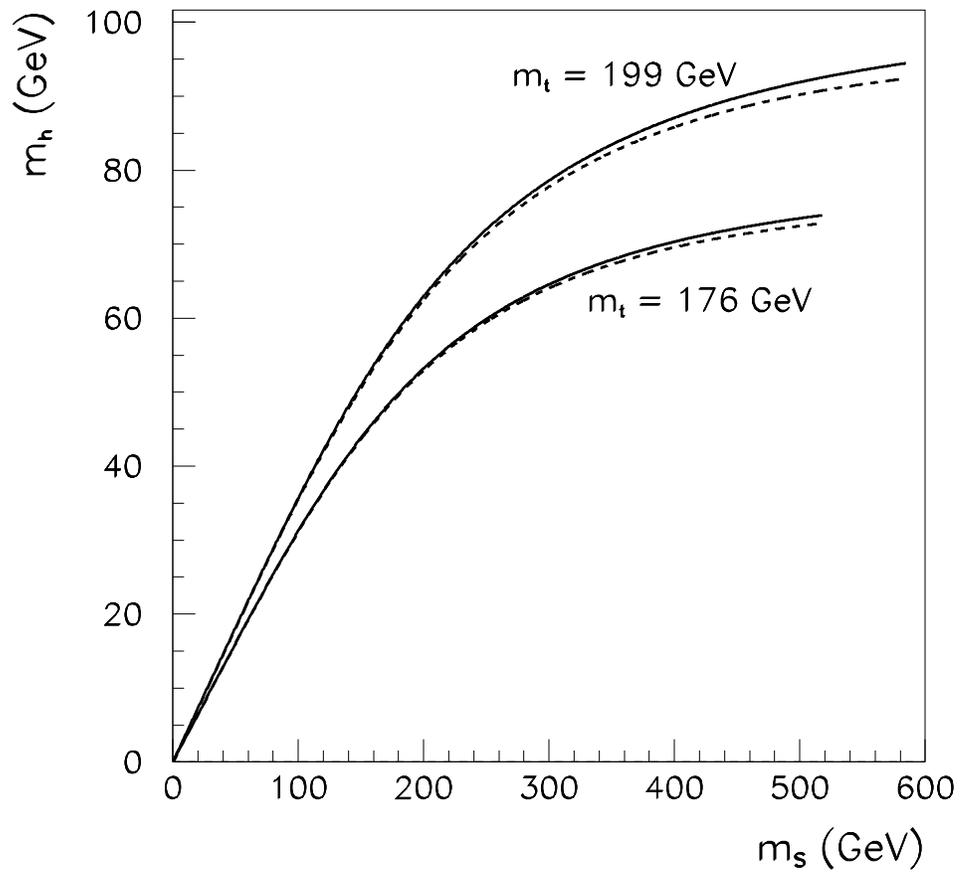
{\bf Fig. 5} \ Higgs boson mass $m_h$ as a function of $m_s$ for
$m_t = 176$ GeV and
$m_t = 199$ GeV. The curves are plotted
for $\Lambda = 1$ TeV (dashed lines) and $\Lambda \rightarrow \infty$ (solid
lines), which indicate the weak dependence of the Higgs mass on the cut off.
The maximal values for $m_h$ correspond to $\tan \beta = 1$.
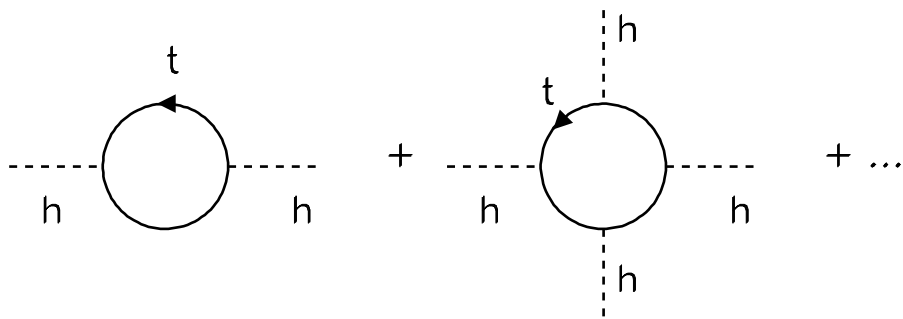
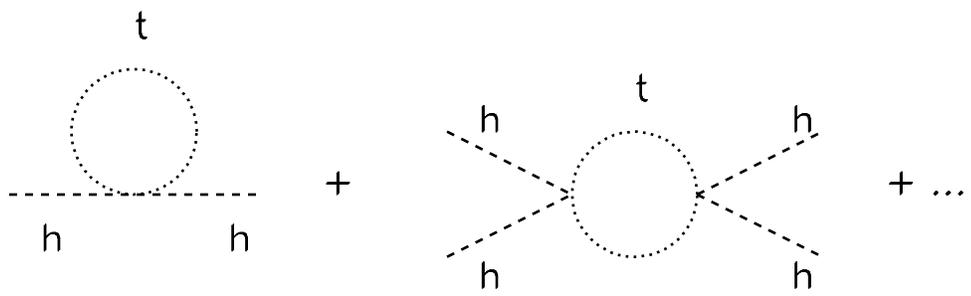
\end{document}